\documentclass[conference,10pt,a4paper]{IEEEtran}

\usepackage{algorithmic,algorithm,graphicx,booktabs,multirow,amssymb,url,cite}
                                                                                                                                                                   
\ifx\pdfoutput\undefined
  \usepackage[pdftex]{graphicx}
\else
  \usepackage{graphicx}
\fi

\def\S{{\cal S}}

\def\A{{\cal A}}
\def\D{{\cal D}}
\def\P{{\cal P}}

\begin{document}

\title{A performance study of anomaly detection using entropy method}

\author{\IEEEauthorblockN{A.A. Waskita\IEEEauthorrefmark{1}\IEEEauthorrefmark{2}, 
H. Suhartanto\IEEEauthorrefmark{2}, L.T. Handoko\IEEEauthorrefmark{4}\IEEEauthorrefmark{5}}
\IEEEauthorblockA{\IEEEauthorrefmark{1}Center for Technology and Safety of Nuclear Reactor, National Nuclear Energy Agency, \\
Kawasan Puspiptek Serpong, Tangerang 15310, Indonesia\\
Email : adhyaksa@batan.go.id}
\IEEEauthorblockA{\IEEEauthorrefmark{2}Faculty of Computer Science, University of
Indonesia, \\
Kampus UI Depok, Depok 16424, Indonesia\\
Email : heru@cs.ui.ac.id}
\IEEEauthorblockA{\IEEEauthorrefmark{4}Group for Theoretical and Computational Physics,
Research Center for Physics, Indonesian Institute of Sciences, \\
Kawasan Puspiptek Serpong, Tangerang 15310, Indonesia\\
Email: laksana.tri.handoko@lipi.go.id}
\IEEEauthorblockA{\IEEEauthorrefmark{5}Department of Physics, University of Indonesia,\\
Kampus UI Depok, Depok 16424, Indonesia\\
Email: handoko@fisika.ui.ac.id}
}

\maketitle

\begin{abstract}
An experiment to study the entropy method for an anomaly detection system has been performed. The study has been 
conducted using real data generated from the distributed sensor networks at the Intel Berkeley Research Laboratory. The experimental results were compared with the elliptical method and has been analyzed in two dimensional data sets acquired from temperature and humidity sensors across 52 micro controllers. Using the binary classification to determine the upper and lower boundaries for each series of sensors, it has been shown that the entropy method are able to detect more number of out ranging sensor nodes than the elliptical methods. It can be argued that the better result was mainly due to the lack of elliptical approach which is requiring certain correlation between two sensor series, while in the entropy approach each sensor series is treated independently. This is very important in the current case where both sensor series are not correlated each other.
%
\end{abstract}


\section{Introduction}
\label{sec:intro}

Detecting anomaly, especially in a safety critical system is very important to mitigate any system failures 
in the near future \cite{BORING}. In some systems, such failures could lead to the tremendous environmental 
disasters. Therefore, those systems are always equipped with robust monitoring system based 
on the eiher wireless or wired sensor network (WSN). The network should involve various types and ranges of sensors which 
transmit the acquired data to the central unit. In some cases, the sensors are embedded in the nearby cascade controller 
prior to the main unit relatively far away from the monitoring node in the field. In a more complex system, it could 
consist of several tiers from the main system till the end monitoring nodes. 

Some examples of the systems with tremendous environmental influences are the so-called landslide early warning system 
(LEWS) involving micro-electromechanical system (MEMS) based sensors, fiber optic strain sensing and GPS tracking system 
to monitor the ground motion related to earthquakes or volcanic activities \cite{hanto,slews2}; the forest fire detection and 
monitoring system \cite{fireforest}.

Some techniques to detect the anomalies have originally been developed for cyber security, in particular to 
mitigate the cyber attacks. For instance, the intrusion detection system (IDS) or intrusion prevention system (IPS) 
was worked out by \cite{Kavitha,Fragkiadakis,Lin,infsci}. In term of cyber security, those methods are complement 
to the signature approaches. It should be noted that the signature based IDS performs better in detecting the well 
known patterns of intrusion, while the anomaly based ones suits for the unknown patterns \cite{Elhag}.

\begin{figure*}[t]
  \centering
  \includegraphics[scale=.55]{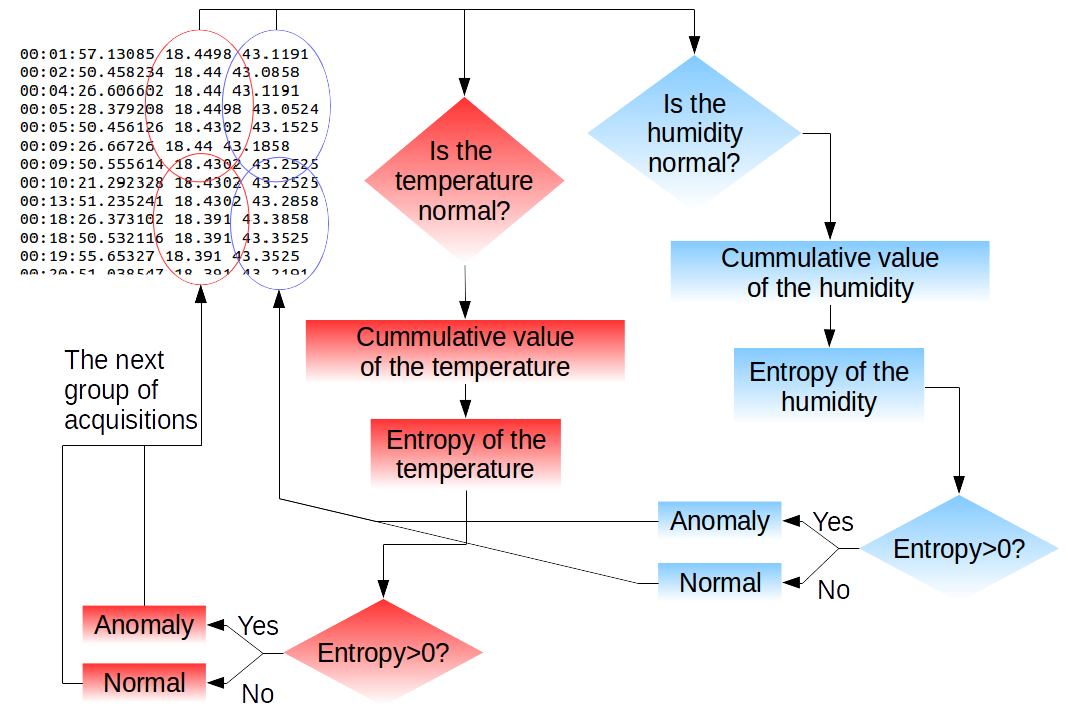}
  \caption{Flowchart of the entropy method calculation used in the present paper \cite{arya3}.}
  \label{fig:flowchart}
\end{figure*}

In contrast with the unpredictable "pattern" in cyber attacks which require the training procedure to define the 
"normal" patterns, in most cases any systems under monitoring through WSN have 
been constructed based on the pre-defined rules or model with certain parameter sets. These parameter sets consequently 
govern the allowed ranges of all sensor nodes within the system. However, the model is perfect under a presumption that 
all sensor and controlling nodes are working well without any failures. Concerning any potential failures during the operation, 
it is considerable to put the so-called early anomaly detection system (EADS) prior to the main processing unit. The 
EADS should lightweight, and not overburden the whole system. It is not necessarily accurate, but it should be able to 
provide, at least preliminary, information of any partial failures in advance. 
Actually in our previous works, the anomaly detection system has also been investigated using the statistical approaches. 
In the approaches a kind of interactions among single or cluster sensor nodes within the system has been modeled through 
the weighted "relationships" among the nodes \cite{arya,arya2}. Unfortunately, the model is quite exhaustive and requires 
huge computing power. 

From this point of view, some approaches based on the "previous" pattern as adopted in the cyber security might not be 
appropriate. It would be better to set up more deterministic approaches like the entropy method \cite{arya3}. The paper 
attempts to apply the entropy based method for the EADS in sensor network. The sensor nodes could be homogeneous or 
hybrid with various characteristics without assuming any interactions among them. The method only measures the level 
of irregularities in the system based on the predefined allowed ranges of each node following its specifications. 
The irregularities at certain degrees within a cluster or the whole system are interpreted as anomalies. As already 
argued in some other previous works \cite{JianQi} and references therein, the entropy based method requires light 
computing power and fast enough for anomaly detection. These natures are suitable for our purpose in the present case.

The paper is organized as follows. After this section, the entropy method is briefly explained in Sec. \ref{sec:method}. 
Sec. \ref{sec:experiment} deals with the experiment using the real data set, and followed with discussion on the 
comparison with the elliptical methods in the previous work by Rajasegarar et.al.. The paper is ended with the summary.

\section{Entropy method}
\label{sec:method}

Following the seminal work of Shannon \cite{Shannon}, the entropy is defined as the level of irregulaties occur, or 
in another word a measure of disorder in a system under consideration. It can be calculated using the master formula 
\cite{arya3},
\begin{equation}
  H = -\sum_{k=1}^{K} p_{k} \log p_k \, ,
  \label{eq:entropy}
\end{equation}
where, 
\begin{equation}
  p_k = \frac{a_k}{\sum_{j=1}^{K} a_j} \, ,
  \label{eq:probabilitas}
\end{equation}
is the elements of probability $\P = \left\{ p_1, p_2, \cdots, p_K \right\}$ of $\D_k$. $\D_k$ is the elements 
of accumulated state set, $\S_A = \left\{ \D_1, \D_2, \cdots, \D_K \right\}$ with $K \leqslant M$, and it is 
composed of all non-repetitive states in $\S$. On the other hand, $a_k$ is the 
elements of $A = \left\{ a_1, a_2, \cdots, a_K \right\}$ which is representing the number of repetitions of $\D_k$. 


Now, these procedure can be applied to investigate the real data and to perform a comparison with another methods.

\section{Experiment}
\label{sec:experiment}

\begin{figure}
  \centering
  \includegraphics[scale=.32]{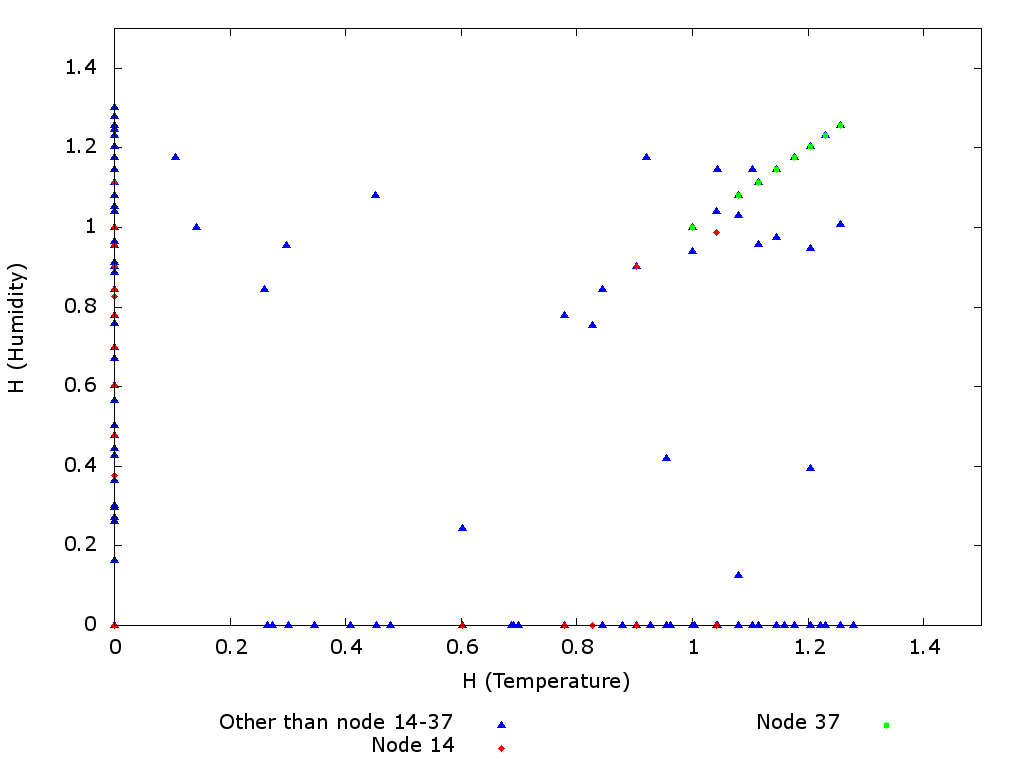}
  \caption{The results of entropies for the series of temperature and humidity sensors. The green, red and blue triangulars are representing the values of the 37th, 14th and remaining nodes. }
  \label{fig:entropyResult}
\end{figure}

The experiment was conducted by taking the real data from the Intel Berkeley Research Laboratory (IBRL) data set 
during the acquisition period of March 1$^\mathrm{st}$, 2004 from 00:00 to 03:59. This period was choosen 
following the work by Rajasegarar et.al \cite{Rajasegarar2007,Rajasegarar2009}. Only temperature and humidity 
sensors were taken into consideration over 54 MICA2DOT microprocessors, where each of them actually consisted of 4 sensor nodes: temperature, humidity, light and voltage. 

As already mentioned in the preceeding section, the entropy method itself does not requires the correlation between 
temperature and humidity sensors. Therefore, one can determine independently the normal ranges for each sensor series 
using its manufacture specifications, and also took into account the fact that all data from the 37th node and a part 
of the 14th node were considered anomaly. Hence, the normal boundary condition $x_b$'s in the current experiment were set to be the following,
\begin{itemize}
  \item Temperature: $15.55^{\circ}$C $< x_b < 18.00^{\circ}$C
  \item Humidity: $42.25\% < x_b < 45.80\%$
\end{itemize}

Further, each sensor series was divided into a smaller interval of time, namely 10 minutes, to have a set of data being calculated using Eqs. (\ref{eq:entropy}) and (\ref{eq:probabilitas}). This procedure was taken to enable the entropy analysis at the node level. Further calculation is illustrated in Fig. \ref{fig:flowchart}. Each acquisition in the 10 minutes interval should be evaluated against the normal boundary condition to determine whether the data acquisition is normal or not. The combination of normal-abnormal of the 10 minutes interval data acquisition establish a cummulative value.

The following is the illustration for determining the entropy from the experiment in Fig. \ref{fig:flowchart}. For the first 10 minutes interval of the figure, there are 7 data acquisition for temperature and humidity parameters. At this interval, all temperature obtained exceeds the normal boundary. Based on the \cite{arya3}, the value of $\A$ for all of them is $1$ and construct the cummulative value array of $\S_A$ as $\{1,2,3,4,5,6,7\}$. With different cummulative value consist of single element, the accumulated state array of $A$ is $\{1,1,1,1,1,1,1\}$. Then, each of the $\S_A$ element has the probability value of $\frac{1}{7}$. This leads to $H = 0.85$. On the other hand, all of the first humidity acquisition data from Fig. \ref{fig:flowchart} are inside the boundary then the value of $\A$ for all of them is $0$ and construct the cummulative value array of $\S_A$ as $\{0,0,0,0,0,0,0\}$. There is only one cummulative value $\left( A=\{7\}\right) $ that produce the probability value into $1$. This leads to $H = 0$.

The algorithm \ref{alg:flowchart} describes the step-by-step of the procedure illustrated in Fig. \ref{fig:flowchart}

\begin{algorithm}
\caption{Evaluate the interaction}
\label{alg:flowchart}
\begin{algorithmic}[1] 
\REQUIRE $x_1,\ldots,x_n$
\COMMENT {The data acquisitions from a sensor in a 10 minutes time interval, the number of $n$ as much as the data captured}
\REQUIRE $x_{b(1)}, x_{b(2)}$
\COMMENT {A normal boundary for a physical parameter, $x_{b(1)}=$lower boundary, $x_{b(2)}=$upper boundary,}
\REQUIRE {$\S_A$}
\COMMENT {An array with $n$ elements of cummulative value}
\REQUIRE {$A$}
\COMMENT {An array listed a number of different cummulative values produced. If only one cummulative value exist, its value should equal to the number of data captured from the sensor in a certain 10 minutes time interval}
\REQUIRE {$\P$}
\COMMENT {An array of the probability for each different cummulative value}
\FOR {$i=1 \to n$}
\IF {$x_{b(1)} \leq x_i \leq x_{b(2)}$} 
\STATE {$\S_{A(i)}=0+\S_{A(i-1)}$}
\ELSE
\STATE {$\S_{A(i)}=1+\S_{A(i-1)}$}
\ENDIF
\ENDFOR
\STATE {construct the array of $A$}
\COMMENT {Its element is the number of different cummulative value occured ($K$)}
\STATE {calculate the array of $\P$}
\COMMENT {Its element is a probability of different cummulative value occured in a certain 10 minutes time interval}
\STATE {$H = -\sum_{k=1}^{K} p_{k} \log p_k $}
\end{algorithmic}
\end{algorithm}

One should note that the 5th and 15th nodes were discarded since the data were missing in the data set. The calculated results for each sensor series are plotted in Fig. \ref{fig:entropyResult}. 

In the next section, the result is be compared with another methods done in some previous works. 

\section{Discussion and summary}
\label{sec:summary}

\begin{figure}
  \centering
  \includegraphics[scale=.33]{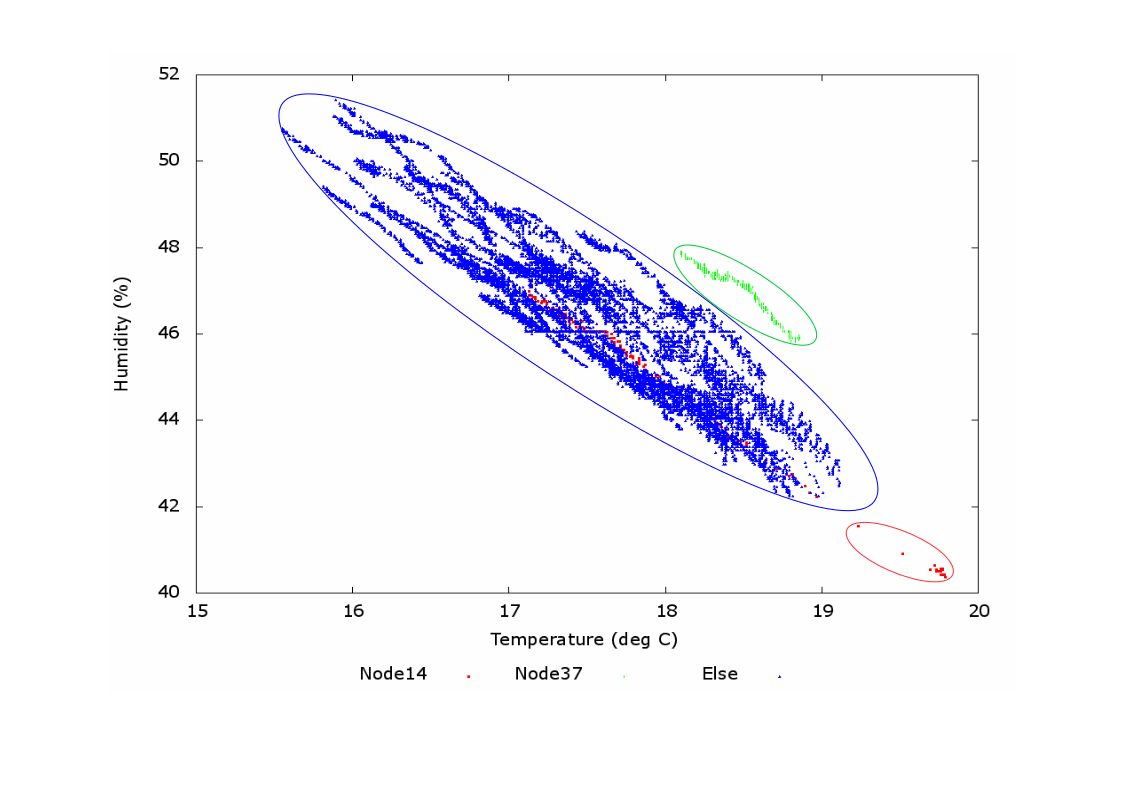}
  \caption{Elliptical method in constructing a normal boundary condition proposed by Rajasegarar et.al \cite{Rajasegarar2009} with the color represents data from the same nodes as depicted in Fig. \ref{fig:entropyResult}.}
  \label{fig:rajasegarar1a}
\end{figure}

The result using entropy method in Fig. \ref{fig:entropyResult} shows the anomalies scattered over the area, while 
the normal data are on both horizontal and vertical axis. In particular, one should notice that all data coming from 
the 37th node (green triangulars) are completely outranged, while only partial part of the 14th node (red triangulars) are 
recognized as anomalies as expected. On the other hand, the entropy has successfully detected the data anomalies coming from 
various nodes. 

One can compare the current result with the previous ones done by Rajasegarar et.al \cite{Rajasegarar2009} using the 
elliptical method. The method is based on the elliptical curves to determine the "normal" region over the data as 
illustrated in Fig. \ref{fig:rajasegarar1a}. In the figure, the $90\%$ CL (confidence level) curve is shown by the 
blue curve generated from the correlation between the data from temperature and humidity sensor nodes. Then, one can 
easily describe more curves with lower CLs to exclude the should-be outlier data. However it is not trivial to fit 
the curve to accomodate the allowed and anomalous regions. 

More detailed study was conducted in the paper by Moshtaghi et.al. using the fractional elliptical method. It also dealt 
with the same data set, but different period of time, to accomodate the inaccuracies in Rajasegarar et.al. \cite{Moshtaghi2009}. 
The fractional elliptical method is able to detect better the outlier data in between the curves. Unfortunately we cannot 
provide the one by one direct comparison due to the different period of data set.

Finally, the present paper has shown the result of EADS using the entropy method, and its comparison with the previous results 
using the elliptical method. The comparison has been conducted in two dimensional space based on the entropies calculated 
from the data series of temperature and humidity nodes. Each value of entropies have been calculated using the data set of 
10 minutes interval along the whole period under consideration. It is argued that the entropy method is able to detect the scattered 
anomalies across the space, regardless its pattern in contrast with, for instance, the elliptical method. 

\section*{Acknowledgments}

AAW thanks the Indonesian Ministry of Research and Technology for financial support, and the Group for Theoretical 
and Computational Physics, Research Center for Physics LIPI for warm hospitality during the work. LTH thanks to 
the Abdus Salam ICTP for hospitality when the initial part of this work was done. LTH is funded by Riset Unggulan LIPI in 
fiscal year 2016 under Contract no. 11.04/SK/KPPI/II/2016.

\bibliographystyle{IEEEtran}
\bibliography{ic3ina2016}
\end{document}